\documentclass[epj]{webofc}
\usepackage[varg]{txfonts}   
\wocname{EPJ Web of Conferences}
\woctitle{CONF12}
\usepackage[utf8]{inputenc}
\usepackage[english]{babel}
\usepackage{amsmath}
\usepackage{amsfonts}
\usepackage{amssymb}
\usepackage{graphicx}
\usepackage{bm}
\usepackage{braket} 
\usepackage[usenames,dvipsnames,svgnames]{xcolor}
\usepackage[colorlinks=true,
            linkcolor=red,
            urlcolor=gray,
            citecolor=blue]{hyperref}
            
\usepackage{tikz}
\usepackage{pgfplots}
\usetikzlibrary{arrows,calc,external}
\DeclareMathOperator{\Imag}{\mathrm{Im}}

\DeclareMathOperator{\Tr}{\mathrm{Tr}}

\newcommand{\mD}{\mathcal{D}}
\newcommand{\mF}{\mathcal{F}}
\newcommand{\mG}{\mathcal{G}}

\newcommand{\mL}{\mathcal{L}}

\newcommand{\mO}{\mathcal{O}}
\newcommand{\mQ}{\mathcal{Q}}

\begin{document}

\selectlanguage{english}
\title{Coupling of $t\bar t$ and $\gamma\gamma$ with a strongly interacting Electroweak Symmetry Breaking Sector}

\author{Rafael L.~Delgado\inst{1}\fnsep\thanks{\email{rdelgadol@ucm.es}. Contributions to the Procs. of the XIIth Quark Confinement and the Hadron Spectrum, Thessaloniki, Greece, August 2016.}
}

\institute{Departamento de F\'isica Te\'orica I, Universidad Complutense de Madrid, E-28040 Madrid, Spain.\\On leave at the Stanford Linear Accelerator Center (SLAC),\\2575 Sand Hill Road, Menlo Park, CA 94025, USA.
}

\abstract{%
We report the coupling of an external $\gamma\gamma$ or $t\bar t$ state to a strongly interacting EWSBS satisfying unitarity. We exploit perturbation theory for those coupling of the external state, whereas the EWSBS is taken as strongly interacting. We use a modified version of the IAM unitarization procedure to model such a strongly interacting regime. The matrix elements $V_LV_L\to V_LV_L$, $V_LV_L\leftrightarrow hh$, $hh\to hh$, $V_LV_L\leftrightarrow\{\gamma\gamma,t\bar t\}$, and $hh\leftrightarrow\{\gamma\gamma,t\bar t\}$ are all computed to NLO in perturbation theory with the Nonlinear Effective Field Theory of the EWSBS, within the Equivalence Theorem.  This allows us to describe resonances of the electroweak sector that may be found at the LHC and their effect on other channels such as $\gamma\gamma$ or $t\bar t$ where they may be discovered.
}

\maketitle

\section{Introduction}
The new boson with a mass of $125\,{\rm GeV}$ found at the LHC~\cite{ATLAS,CMS} would complete the Standard Model (SM) in its minimal version. The experimental collaborations at CERN are looking for deviations of its behaviour from that of the SM Higgs particle~\cite{Aad:2013wqa,Chatrchyan:2013lba}. The electroweak symmetry breaking happens at a scale of $v=246\,{\rm GeV}$. New scalar resonances are constrained up to $600-700\,{\rm GeV}$~\cite{twophotons}. The constraint over new vector bosons reaches even higher in energy~\cite{searches}. Thus, the study of the TeV scale is the goal of LHC Run-II. And it is expectable that new physics related with the Electroweak Symmetry Breaking Sector (EWSBS), if it exists, will be found at this scale.

To model the behaviour of a generic EWSBS with a theory which only includes the particles already known to exist there (the new Higgslike boson $h$ and the longitudinal components of gauge bosons $W_L^\pm$, $Z_L$), the so-called Higgs Effective Field Theory (HEFT) is used~\cite{Alonso:2015fsp,Alonso:2012px,Contino:2013kra,Gavela:2014uta,Gavela:2014vra,Buchalla:2015qju,Buchalla:2013rka,Feruglio:1992wf,Contino:2010mh,Bagger:1993zf,Koulovassilopoulos:1993pw,Burgess:1999ha,Wang:2006im,Grinstein:2007iv,Brivio:2014pfa,Rauch:2016pai}. A summary of these efforts can be found on the 4th CERN Yellow Report~\cite{YellowReport}. This model is based on the old (Higgsless) Electroweak Chiral Lagrangian (ECL)~\cite{Appelquist}, which is also inspired by Chiral Perturbation Theory (ChPt) for hadron physics~\cite{ChPT}.

In our Refs.~\cite{Delgado:2013loa,Delgado:2013hxa,Delgado:2014dxa,Delgado:2014jda,Delgado:2015kxa}, we simplify the study of the HEFT by means of the Equivalence Theorem~\cite{ET}, which is valid for
\begin{equation}\label{eq:valid:range}
 s\gg M_h^2,M_W^2,M_Z^2\sim (100\,{\rm GeV})^2.
\end{equation}
Since the possible resonance region is above $500\,{\rm GeV}$, this approximation is safe, and allows us to identify the longitudinal $W_L^\pm$, $Z_L$ with the pseudo-Goldstone bosons of symmetry breaking $\omega^a$ ($a=1,2,3$) in their scattering amplitudes. For instance,
\begin{equation}
  T(W_L^i W_L^j\to W_L^k W_L^l) = T(\omega^i\omega^j\to\omega^k\omega^l) + \mO\left(\frac{M_W}{\sqrt{s}}\right) .
\end{equation}
We have found~\cite{Delgado:2013loa,Delgado:2013hxa,Delgado:2014dxa,Delgado:2014jda,Delgado:2015kxa} that, for any parameter choice separating from the SM, the theory becomes strongly interacting at sufficiently high energy, and resonances may appear. Dispersion relations (as unitarization procedures) are used to deal with this non-perturbative regime. This also happens for the ChPT theory for hadron physics~\cite{Truong:1988zp,Dobado:1996ps}. In Ref.~\cite{Dobado:1989gr,GomezNicola:2001as}, unitarization procedures are tested for the hadron case of ChPT.

\begin{figure}\centering
\tikzsetnextfilename{chiral_ttbar}
\begin{tikzpicture}[
        scale=.8,
        axis/.style={very thick, ->, >=stealth', line join=miter},
        dot_red/.style={circle,fill=red,minimum size=4pt,inner sep=0pt,
            outer sep=-1pt},
        dot_blue/.style={circle,fill=blue,minimum size=4pt,inner sep=0pt,
            outer sep=-1pt},
    ]

\draw[axis,<->] (5,0) node(xline)[right] {$M_t$} -| (0,5) node(yline)[above] {$s$};   

\draw (1, 0.1) -- (1, -0.1) node [below] {$M_t\phantom{^1}$};
\foreach \x in {2,...,4}
    \draw (\x, 0.1) -- (\x, -0.1) node [below] {$M_t^\x$};

\draw (0.1, 1) -- (-0.1, 1) node [left] {$s^{1/2}$};
\draw (0.1, 2) -- (-0.1, 2) node [left] {$s\phantom{^{2/2}}$};   
\draw (0.1, 3) -- (-0.1, 3) node [left] {$s^{3/2}$};
\draw (0.1, 4) -- (-0.1, 4) node [left] {$s^{2\phantom{/2}}$};

\draw (-0.1,-0.1) node[below left] {$0$};

\node[dot_red,label=above right:LO] at (0,2) {};
\node[dot_red,label=above right:NLO] at (0,4) {};

\node[dot_blue,label=above right:LO] at (1,1) {};
\node[dot_blue,label=above right:NLO] at (1,3) {};

\node[dot_red,label=right:EWSBS alone] at (4,4) {};
\node[dot_blue,label=right:EWSBS+$t\bar{t}$] at (4,3) {};

\end{tikzpicture}
\hspace{.5cm}
\tikzsetnextfilename{chiral_gamma_gamma}
\begin{tikzpicture}[
        scale=.8,
        axis/.style={very thick, ->, >=stealth', line join=miter},
        dot_red/.style={circle,fill=red,minimum size=4pt,inner sep=0pt,
            outer sep=-1pt},
        dot_green/.style={circle,fill=green,minimum size=4pt,inner sep=0pt,
            outer sep=-1pt},
    ]

\draw[axis,<->] (5,0) node(xline)[right] {$\alpha$} -| (0,5) node(yline)[above] {$s$};   

\draw (1, 0.1) -- (1, -0.1) node [below] {$\alpha\phantom{^1}$};
\foreach \x in {2,...,4}
    \draw (\x, 0.1) -- (\x, -0.1) node [below] {$\alpha^\x$};

\draw (0.1, 1) -- (-0.1, 1) node [left] {$s^{1/2}$};
\draw (0.1, 2) -- (-0.1, 2) node [left] {$s\phantom{^{2/2}}$};   
\draw (0.1, 3) -- (-0.1, 3) node [left] {$s^{3/2}$};
\draw (0.1, 4) -- (-0.1, 4) node [left] {$s^{2\phantom{/2}}$};

\draw (-0.1,-0.1) node[below left] {$0$};

\node[dot_red,label=above right:LO ] at (0,2) {};
\node[dot_red,label=above right:NLO] at (0,4) {};

\node[dot_green,label=above right:LO ] at (1,0) {};
\node[dot_green,label=above right:NLO] at (1,2) {};

\node[dot_red,label=right:EWSBS alone] at (3,4) {};
\node[dot_green,label=right:EWSBS+$\gamma\gamma$] at (3,3) {};
\end{tikzpicture}
\caption{Left: chiral $(M_t,\sqrt{s})$ counting for coupling with $t\bar t$. Right: chiral $(\alpha,s)$ counting for coupling with $\gamma\gamma$. Note that the HEFT is perturbative in $M_t/v$ and $\alpha_{\rm EM}$, respectively, but requires unitarization in $s/(4\pi v)^2$ (all orders along the OY axis included) to reach the resonance region.\label{fig:counting}}
\end{figure}
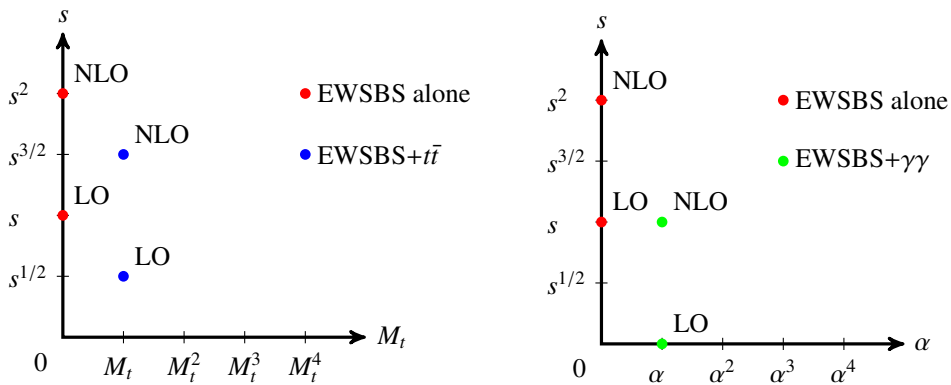

In this work, based on our recent works~\cite{Delgado:2016rtd,Castillo:2016erh}, we couple $\gamma\gamma$ and $t\bar t$ states to the unitarized scattering amplitudes of our refs.~\cite{Delgado:2014dxa,Delgado:2015kxa}. The couplings between the EWSBS and $\{t\bar t,\gamma\gamma\}$ states are perturbative in $M_t/v$ (negligible if compared with $\sqrt{s}/v$) and $\alpha_{\rm EM}$, respectively. However, the EWSBS itself is strongly interacting (see Fig.~\ref{fig:counting}). For the $t\bar t$ case, account must be taken  that we are in the regime $M_t^2/v^2\ll \sqrt{s}M_t/v^2\ll s/v^2$. This allows us to neglect diagrams like the 1-loop top mass renormalization, since they contribute to the scattering amplitudes with higher orders of $\sqrt{s}/v$. Note also that, in the high energy limit $s\ll M_{Z,W}^2\sim M_h^2$ where the Equivalence Theorem can be applied, we can take $M_W^2=M_Z^2=M_h^2=0$ consistently.

Note that we compute $\{\omega,\omega\}\leftrightarrow\{\omega\omega,hh,\gamma\gamma,t \bar t\}$. $\gamma\gamma$ and $t\bar t$ can both appear as initial and final states. Of course, $\gamma\gamma$ and $t\bar t$ as final states are being tested at the LHC experiments, and physics of two photon decays has been pursued since the dawn of particle physics~\cite{Yang:1950rg}. But we should not forget the physics of $\gamma\gamma\to\omega\omega$, since the photon can be a parton of the proton~\cite{Manohar:2016nzj} or the electron in $pp$ and $e^-e^+$ colliders, respectively. The CMS collaboration~\cite{Khachatryan:2016mud} is currently setting bounds to anomalous quartic gauge couplings by analysing $\gamma\gamma\to W^+ W^-$. Photon colliders driven by lepton beams due to Compton backscattering could also become a future application of this work~\cite{Telnov:2016lzw,Gronberg:2014yfa}.

\section{The Electroweak Chiral Lagrangian}
First we quote the HEFT Lagrangian~\cite{Delgado:2014jda,Delgado:2015kxa,Castillo:2016erh} which includes only the low-energy dynamics of the quark sector and the four light modes: three would-be Goldstone Bosons $\omega^a$ (WBGBs) and the Higgs-like particle $h$. A global symmetry breaking pattern $SU(2)_L\times SU(2)_R\to SU(2)_C$ is considered.
\begin{equation}\label{EW_Lagrangian}
   \mL = \frac{v^2}{4}\mF(h/v)\Tr[(D_\mu U)^\dagger D^\mu U] 
       + \frac{1}{2}\partial_\mu h\partial^\mu h
       - V(h)
       + i\bar{Q}\partial Q - v\mG(h/v)[\bar{Q}'_LUH_QQ'_R + {\rm h.c.}] .
\end{equation}
The $U(x)\in SU(2)_C$ isospin can be parametrized by using the so-called spherical parametrization~\cite{Delgado:2015kxa} by means of
\begin{equation}
   U=\sqrt{1-\frac{\omega^2}{v^2}}+i\frac{\omega^i}{v}\tau_i .
\end{equation}
Note that $i=1,2,3$ in the isospin basis, which is related to the charge one by $\omega^\pm = (\omega^1\mp i\omega^2)/\sqrt{2}$ and $\omega^0=\omega^3$. The $SU(2)_L\times U(1)_Y$ covariant derivative is given by
\begin{equation}
  D_\mu U = \partial_\mu U + i\frac{g}{2}\tau_i U W_\mu^i - i\frac{g'}{2} U \tau_3 B_\mu 
          = i\frac{\partial_\mu\omega^i}{v}\tau_i + i\frac{g}{2}W_{\mu}^i\tau_i - i\frac{g'}{2}B_\mu \tau_3 + \dots,
\end{equation}
where the dots represent terms of higher order in $(\omega^a/v)$.

The Higgs potential is expanded as
\begin{equation}
  V(h/v) = v^4\sum_{n=3}^\infty V_n\left(\frac{h}{v}\right)^n.
\end{equation}
The SM is recovered for $V_3 = M_h^2/2v^2$, $V_4=M_h^2/8v^2$, $V_n=0\;\forall n>4$. Note that these terms are subleading in our approximation (Eq.~\ref{eq:valid:range}), so that we could neglect the whole potential $V$ provided that this behaviour holds by whatever beyond SM theory happens to succeed. This is the case in most models of interest, and is a reasonable hypothesis since the constraints of these couplings have so far been found to be close to the SM values.

By expanding Eq.~(\ref{EW_Lagrangian}) without considering (yet) the Yukawa part, we obtain
\begin{align}\label{ec:expand:Lagrangian}
   \mL =&{} \frac{1}{2}\partial_\mu h\partial^\mu h
       + \frac{1}{2}\mF(h/v)
          (2\partial_\mu\omega^+\partial^\mu\omega^- + \partial_\mu\omega^0\partial^\mu\omega^0)
      \notag\\&
       + \frac{1}{2v^2}\mF(h/v)
          (\partial_\mu\omega^+\omega^- + \omega^+\partial_\mu\omega^-+\omega^0\partial_\mu\omega^0)^2
      \notag\\&
       + ie\mF(h/v)A^\mu(\partial_\mu\omega^+\omega^- - \omega^+\partial_\mu\omega^-)
       + e^2\mF(h/v)A_\mu A^\mu \omega^+\omega^- ,
\end{align}
where $A_\mu = \sin\theta_W W_{\mu,3}+\cos\theta_W B_\mu$. We are not considering couplings with external transverse gauge bosons. Note that these neglected states do not appear in inner loops since this would be a higher order correction in $\alpha_{\rm EM}$.

The Yukawa part of the Lagrangian of Eq.~(\ref{EW_Lagrangian}), once the Yukawa-coupling matrix is diagonalized~\cite{Castillo:2016erh}, can be written as
\begin{multline}
\mL_Y =-\mG\left( h\right) \left[%
    \sqrt{1-\frac{\omega^2}{v^2}}\left( M_t t\bar{t} + M_b\bar{b}b \right)
    +\frac{i\omega^0}{v}\left(M_t \bar{t}\gamma^5 t + M_b\bar{b}\gamma^5 b\right)
 \right.\\\left.
    +i\sqrt{2}\frac{\omega^+}{v}\left( M_b\bar{t}_L b_R - M_t\bar{t}_R b_L\right)
    +i\sqrt{2}\frac{\omega^-}{v}\left( M_t\bar{b}_L t_R - M_b\bar{b}_R t_L\right)
\right] .  \label{LY1G}
\end{multline}
Note that we have considered only couplings with the third quark generation ($M_{t,b}\gg M_{c,s,u,d}\to 0$). This Lagrangian breaks custodial symmetry because of $M_t\gg M_b$.

The $\mF$ and $\mG$ functions of Eq.~(\ref{EW_Lagrangian}) are parametrized as
\begin{align}
   \mF(h/v) &= 1 + 2a\frac{h}{v} + b\frac{h^2}{v^2} + \dots & 
   \mG(h/v) &= 1 + c_1\frac{h}{v} + c_2\frac{h^2}{v^2} + \dots,
\end{align}
so that Eq.~(\ref{LY1G}) can be written as
\begin{equation}
\mL_Y = -\left( 1+c_1\frac{h}{v}+c_2\frac{h^2}{v^2}\right)\left[
      \left(1-\frac{\omega^2}{2v^2}\right) M_t t\bar{t}  
    +\frac{i\omega^0}{v}M_t\bar{t}\gamma^5 t
    -i\sqrt{2}\frac{\omega^+}{v} M_t\bar{t}_R b_L
    +i\sqrt{2}\frac{\omega^-}{v} M_t\bar{b}_L t_R\right].
\end{equation}

Since this is an EFT, in order to renormalize the tree level Lagrangian in Eq.~(\ref{ec:expand:Lagrangian}) at the one-loop level, counterterms of dim.~8 are needed. For the EWSBS (strong) interactions, the minimal set of counterterms is $\{a_4,a_5,d,e,g\}$ (see Refs.~\cite{Delgado:2013hxa,Delgado:2015kxa}),
\begin{multline}
\mL_{4,\rm EWSBS} =
    \frac{4a_4}{v^4}\partial_\mu\omega^i\partial_\nu\omega^i\partial^\mu\omega^j\partial^\nu\omega^j
   +\frac{4a_5}{v^4}\partial_\mu\omega^i\partial^\mu\omega^i\partial_\nu\omega^j\partial^\nu\omega^j \\
   +\frac{2d}{v^4}\partial_\mu h\partial^\mu h\partial_\nu\omega^i\partial^\nu\omega^i
   +\frac{2e}{v^4}\partial_\mu h\partial^\nu h\partial^\mu\omega^i\partial_\nu\omega^i
   + \frac{g}{v^4}\left(\partial_\mu h\partial^\mu h\right)^2
\end{multline}
For the $\gamma\gamma$ coupling~\cite{Delgado:2014jda,Delgado:2016rtd} we consider $\{a_1,a_2,a_3,c_\gamma\}$, though only $a_1+a_2-a_3$ appears and none is strictly needed,
\begin{multline}
\mL_{4,\gamma\gamma} = %
\frac{e^2a_1}{2v^2}A_{\mu\nu}A^{\mu\nu}\left(v^2 - 4\omega^+\omega^-\right) %
 + \frac{2e(a_2-a_3)}{v^2}A_{\mu\nu}\left[%
         i\left(\partial^\nu\omega^+\partial^\mu\omega^- - \partial^\mu\omega^+\partial^\nu\omega^- \right) %
   \right. \\ \left. %
        +eA^\mu\left( \omega^+\partial^\nu\omega^- + \omega^-\partial^\nu\omega^+ \right)
        -eA^\nu\left( \omega^+\partial^\mu\omega^- + \omega^-\partial^\mu\omega^+ \right)
        \right] -\frac{c_{\gamma}}{2}\frac{h}{v}e^2 A_{\mu\nu} A^{\mu\nu} .
\end{multline}
And for the Yukawa coupling with top quarks~\cite{Castillo:2016erh},
\begin{equation}
\mL_{4,t\bar t} = %
    g_t\frac{M_t}{v^4} (\partial_\mu\omega^i\partial^\mu\omega^j) t\bar t
  +g'_t\frac{M_t}{v^4} (\partial_\mu h\partial^\mu h) t\bar t
\end{equation}

\section{Partial waves}
The perturbative $\{\omega^i\omega^j,hh\}\to\{\omega^k\omega^l,hh\}$ scattering amplitudes can be found in our Ref.~\cite{Delgado:2015kxa}. For the representation of the $\omega\omega$ states, we will use the isospin basis $\ket{I,M_I}$ ($I$ is the isospin and $M_I$, its projection). Because of isospin symmetry, scattering amplitudes factorizes in this basis, and their values do not depend on $M_I$. Unitarization procedures (based on dispersion relations) are most easily applicable over a partial wave decomposition~\cite{Delgado:2015kxa} which, for $\{\omega^i\omega^j,hh\}$ states, are computed by means of
\begin{equation}\label{part_waves:EWSBS}
 A_{IJ}(s) = \frac{1}{64\pi}\int_{-1}^1 d(\cos\theta)P_J(\cos\theta)A_I(s,t,u) .
\end{equation}
Since the $hh$ state is an isospin singlet $I=0$ (which couples with $J=0,2$), if $I\neq 0$ there is no mixing with $hh$ channel, and the non-vanishing matrix elements are~\cite{Delgado:2015kxa} $IJ=11,20,22$. A similar expression to Eq.~(\ref{part_waves:EWSBS}) is used for $M_I(\omega^i\omega^j\to hh)$ and $T_I(hh\to hh)$ ($J=0$ in these case).

Note that partial waves of Eq.~(\ref{part_waves:EWSBS}) have a chiral expansion
\begin{equation}
  A_{IJ}(s) = A^{(0)}_{IJ}(s) + A^{(1)}_{IJ}(s)+\dots,
\end{equation}
where $A^{(0)}_{IJ}(s)\sim s$ corresponds to the LO term and $A^{(1)}_{IJ}(s)\sim s^2$, to the NLO computation and counterterms.

For the $\gamma\gamma$ states, according to our Refs.~\cite{Delgado:2014jda,Delgado:2016rtd}, we have 4 polarization states that will be labeled as $\lambda_1\lambda_2\in\{++,+-,-+,--\}$. The equation equivalent to Eq.~(\ref{part_waves:EWSBS}), taking into account the effects of polarization, is
\begin{equation}
  P_{IJ}^{\lambda_1\lambda_2} = \frac{1}{128\pi^2}\sqrt{\frac{4\pi}{2J+1}}\int d\Omega\,T_I^{\lambda_1\lambda_2}(s,\Omega)Y_{J,\Lambda}(\Omega),%
  \;\Lambda=\lambda_1-\lambda_2 .
\end{equation}
Note that only $\ket{I,M_I}=\ket{0,0},\ket{2,0}$ couples with 2-$\gamma$ states, due to electric charge and angular momentum conservation. Indeed, parity conservation forbids 2-$\gamma$ negative parity states coupling with $\{\omega^i\omega^j,hh\}$. Thus, let us introduce the positive parity state $(\ket{+-}+\ket{-+})/\sqrt{2}$. This is the only one that couples with $J=0$ states. Hence, we can define $P_{I0}\equiv (P_{I0}^{++}+P_{I0}^{--})$. For $J=2$, the only non-vanising contributions come from the (positive parity states) $\ket{+-}$ and $\ket{-+}$, so that $P_{I2}\equiv P_{I2}^{+-}=P_{I2}^{-+}$. The perturbative scattering amplitudes themselves be found in Ref.~\cite{Delgado:2014jda}.

Finally, the $t^{\lambda_1}\bar t^{\lambda_2}$ states only couple with $\ket{I=0} = \sum_i\ket{\omega^i\omega^i}/\sqrt{3}$ and $hh$ states~\cite{Castillo:2016erh}. Even more, $\ket{I=0}$ only couples with $\ket{S=1,S_Z=0} = (\ket{+,+}-\ket{-,-})/\sqrt{2}$ $t\bar t$ state. The corresponding partial waves are
\begin{equation}
Q(\omega\omega\to t\bar t) = \frac{\sqrt{2}}{64\pi^2}\int d\Omega\mD_{00}^0(\phi,\theta,-\phi)\mQ (\omega\omega\to t^+\bar{t}^+),
\end{equation}
and a similar expression for $Q(hh\to t\bar t)$.

\section{Unitarization procedures}\label{sec:unit_proced}
As pointed out in Fig.~\ref{fig:counting}, our key assumption will be a strongly interacting EWSBS while couplings with $\gamma\gamma$ and $t\bar t$ states remain perturbative. The EWSBS $\{\omega^i\omega^j,hh\}$ scattering partial waves are unitarized by means of the same unitarization procedures (IAM, N/D, Improved-K matrix) that were exposed on our Refs.~\cite{Delgado:2014dxa,Delgado:2015kxa}.

We have two possibilites when unitarizing the EWSBS~\cite{Delgado:2015kxa}. $\omega^i\omega^j$ states couple with $hh$ if $IJ=00,\,20$ and some of this conditions are verified: $a^2\neq b$, $d\neq 0$ or $e\neq 0$. Otherwise, the $hh$ channel decouples. For instance, the single channel IAM is~\cite{Delgado:2015kxa}
\begin{equation}\label{ec:IAM_EWSBS}
   \tilde{A}^{\rm IAM}_{IJ} = \frac{[A^{(0)}_{IJ}(s)]^2}{A^{(0)}_{IJ}(s)-A^{(1)}_{IJ}(s)}
\end{equation}
Note that if coupling with $hh$ channel happens, the coupled-channel versions of the unitarization procedures are required. Indeed, the strongly interacting regime could be triggered by $\omega\omega\to hh$ even if $\omega\omega\to\omega\omega$ is weak~\cite{Delgado:2014dxa}.

Couplings with $\gamma\gamma$ and $t\bar t$ are \emph{unitarized} by means of a modification of the IAM and N/D procedures that uses as input the unitarized partial waves $\{\omega^i\omega^j,hh\}\to\{\omega^i\omega^j,hh\}$. This modification can be found in our Refs.~\cite{Delgado:2016rtd,Castillo:2016erh}. For the single-channel version of the IAM, it can be used
\begin{equation}\label{ec:PERTURB}
   \tilde{P} = P^{(0)}\frac{\tilde{A}^{\rm IAM}}{A^{(0)}},
\end{equation}
where $\tilde{P}$, in this case, is the unitarized $\omega\omega\to\{\gamma\gamma,t\bar t\}$ amplitude; $P^{(0)}$, the corresponding perturbative one; $\tilde{A}$, the unitarized $A(\omega\omega\to\omega\omega)$; and $A^{(0)}(s)$, the tree level matrix element $A(\omega\omega\to\omega\omega)$. This guarantees that the phase of $\tilde{P}$ coincides with that of $\tilde{A}^{\rm IAM}$. For other channels, see Refs.~\cite{Delgado:2014dxa,Delgado:2015kxa}.

\section{Numerical examples}
\begin{figure}
\includegraphics[width=.33\textwidth]{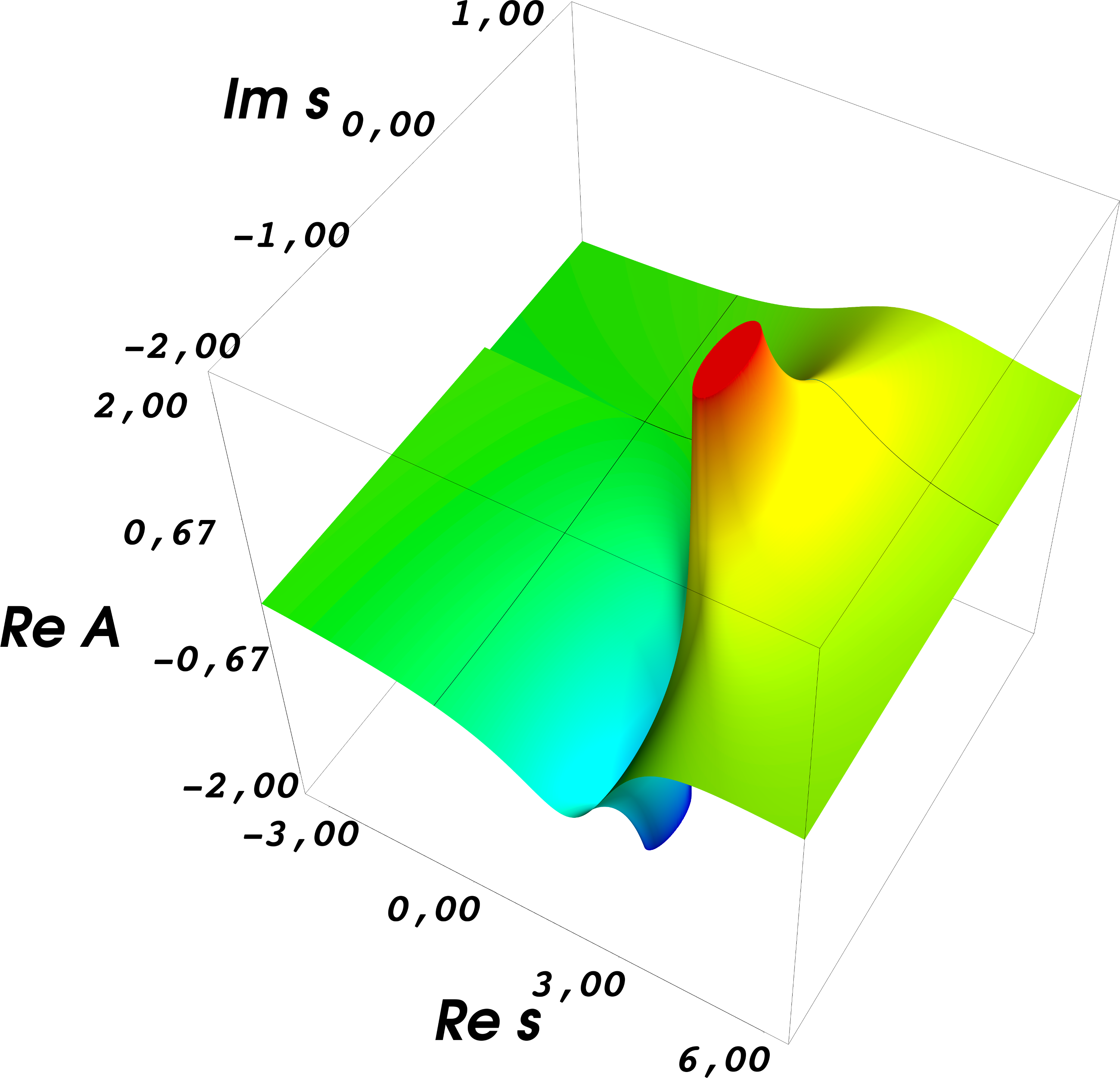}\hfill
\includegraphics[width=.33\textwidth]{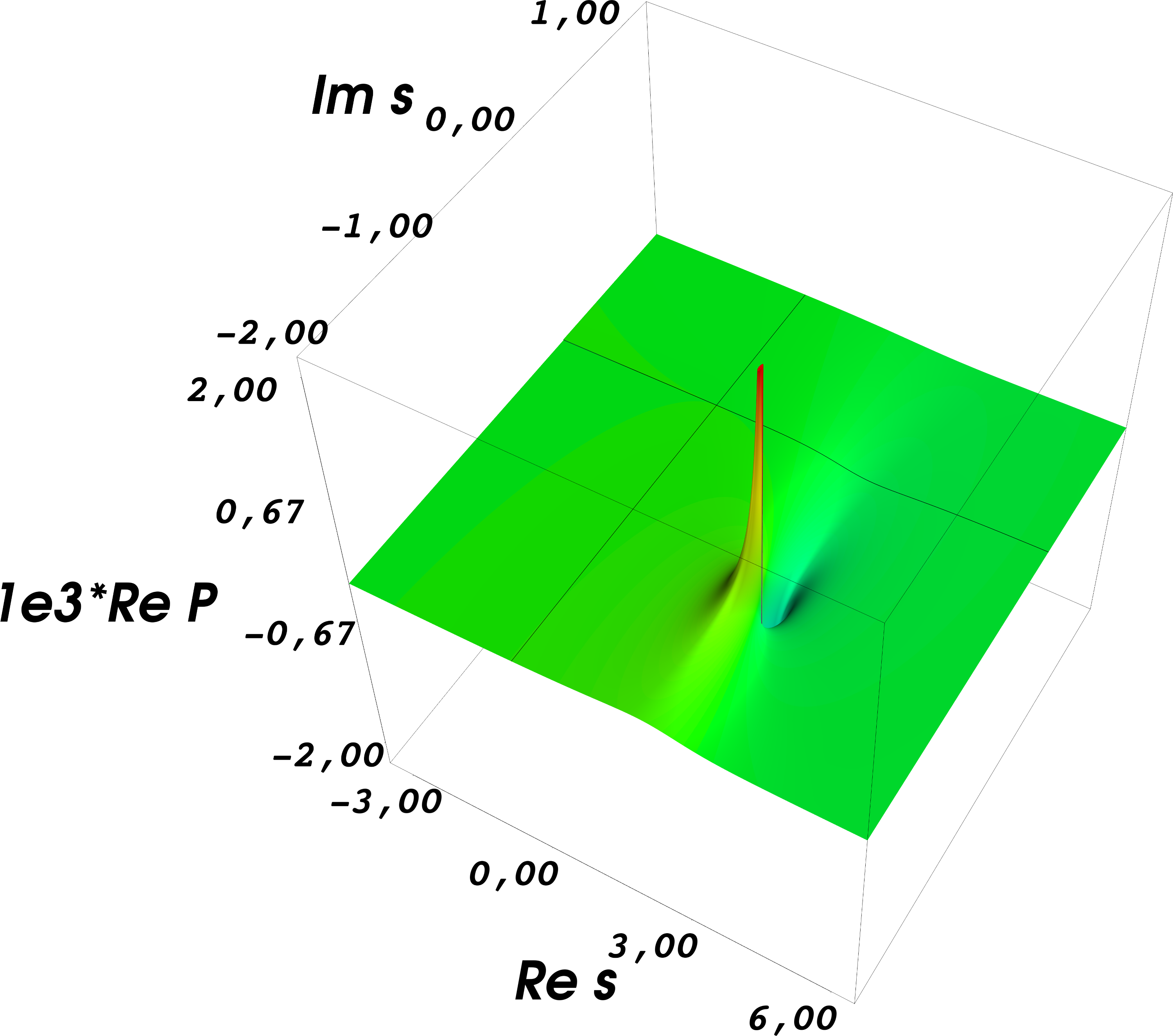}\hfill
\includegraphics[width=.33\textwidth]{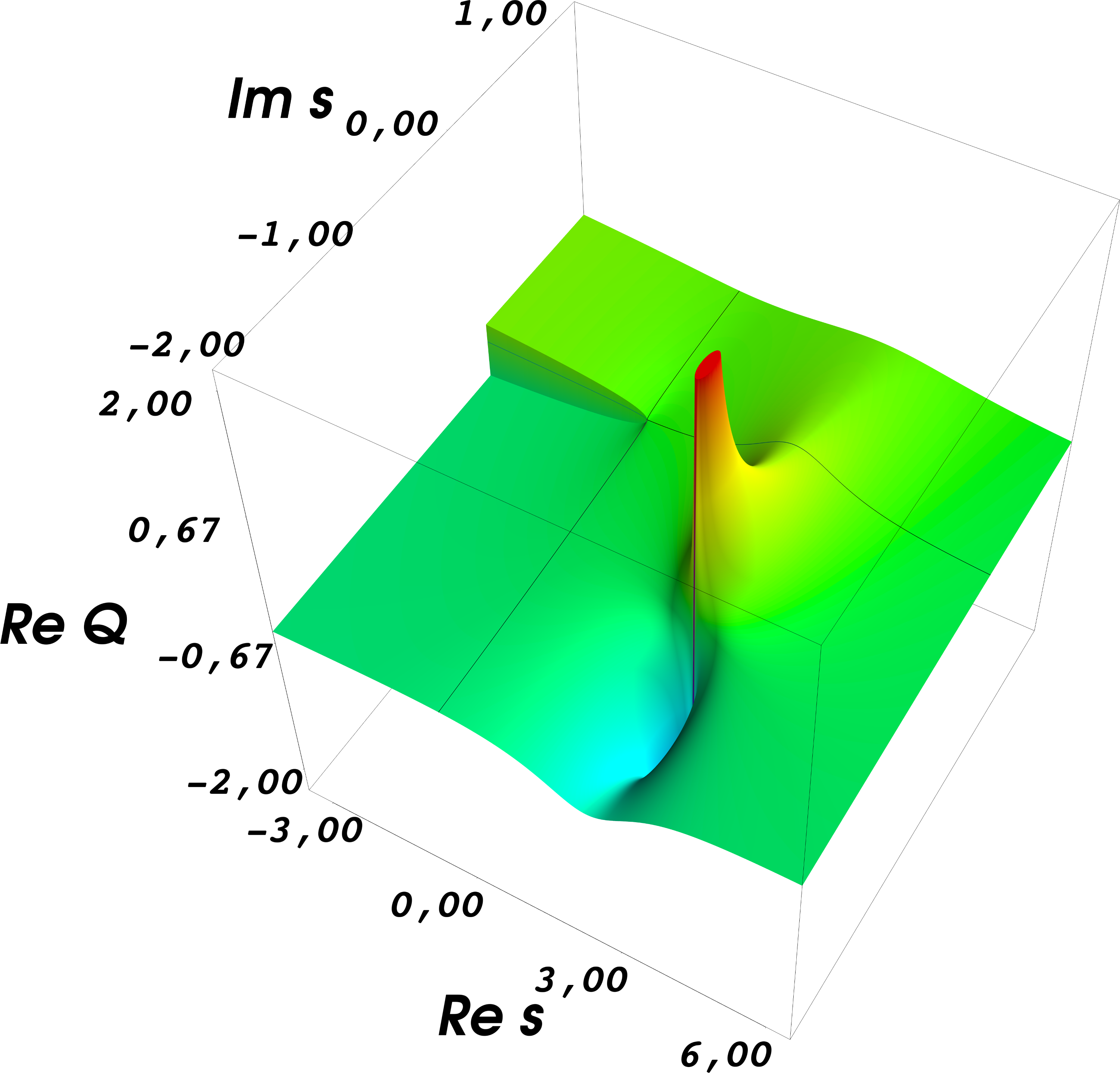}
\caption{From left to right, second Riemann sheets of $\Imag A_{00}^{\rm IAM}$, $10^3\times\Imag P_{00}$ and $\Imag Q$. Parameters: $a^2=b=0.81^2$, $a4=4\times 10^{-4}$, $g=10^{-3}$. All the other NLO parameters set to $0$.}\label{fig:examples1}
\end{figure}
\begin{figure}
\includegraphics[width=.49\textwidth]{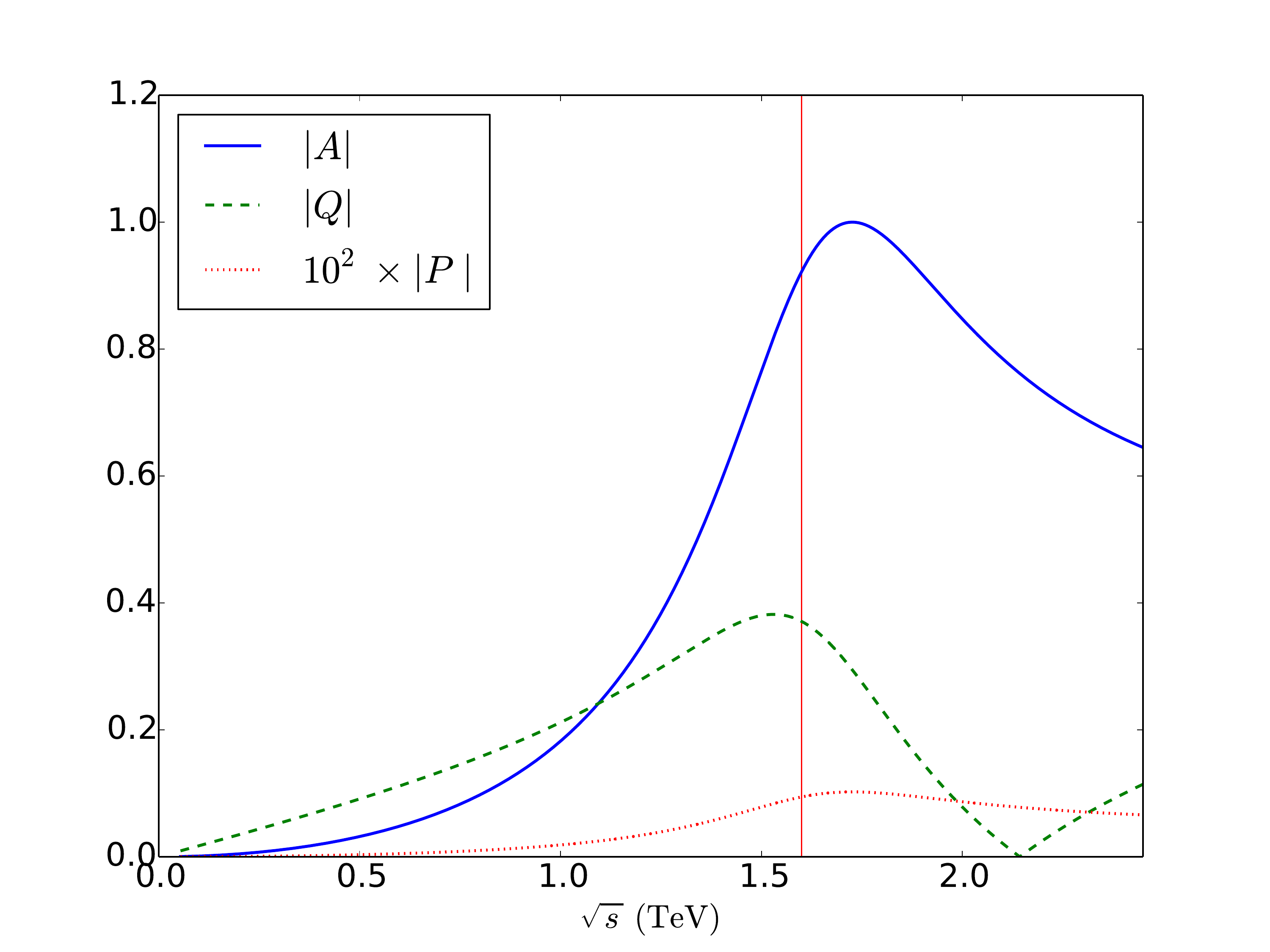}\hfill
\includegraphics[width=.49\textwidth]{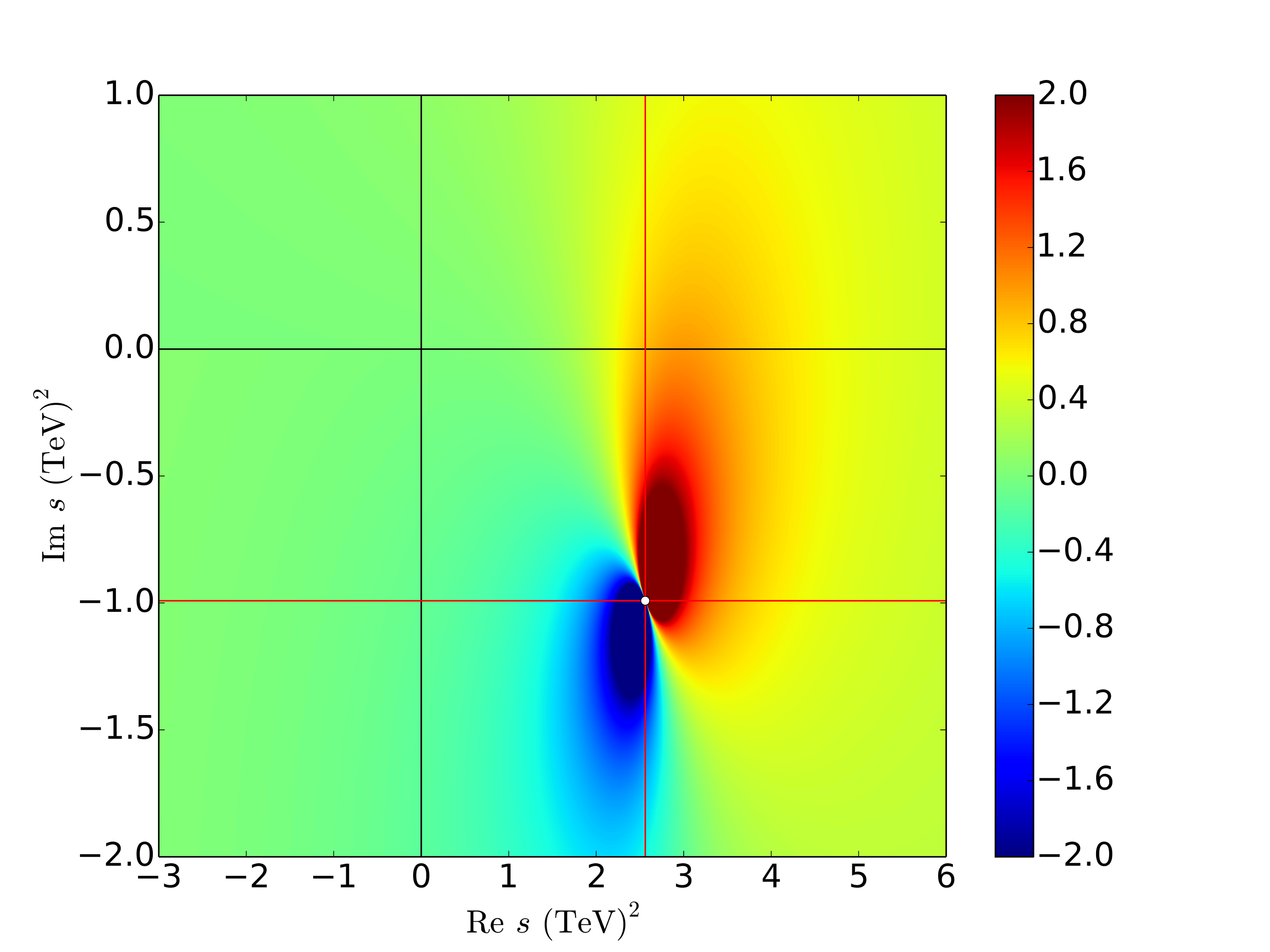}\hfill \\
\includegraphics[width=.49\textwidth]{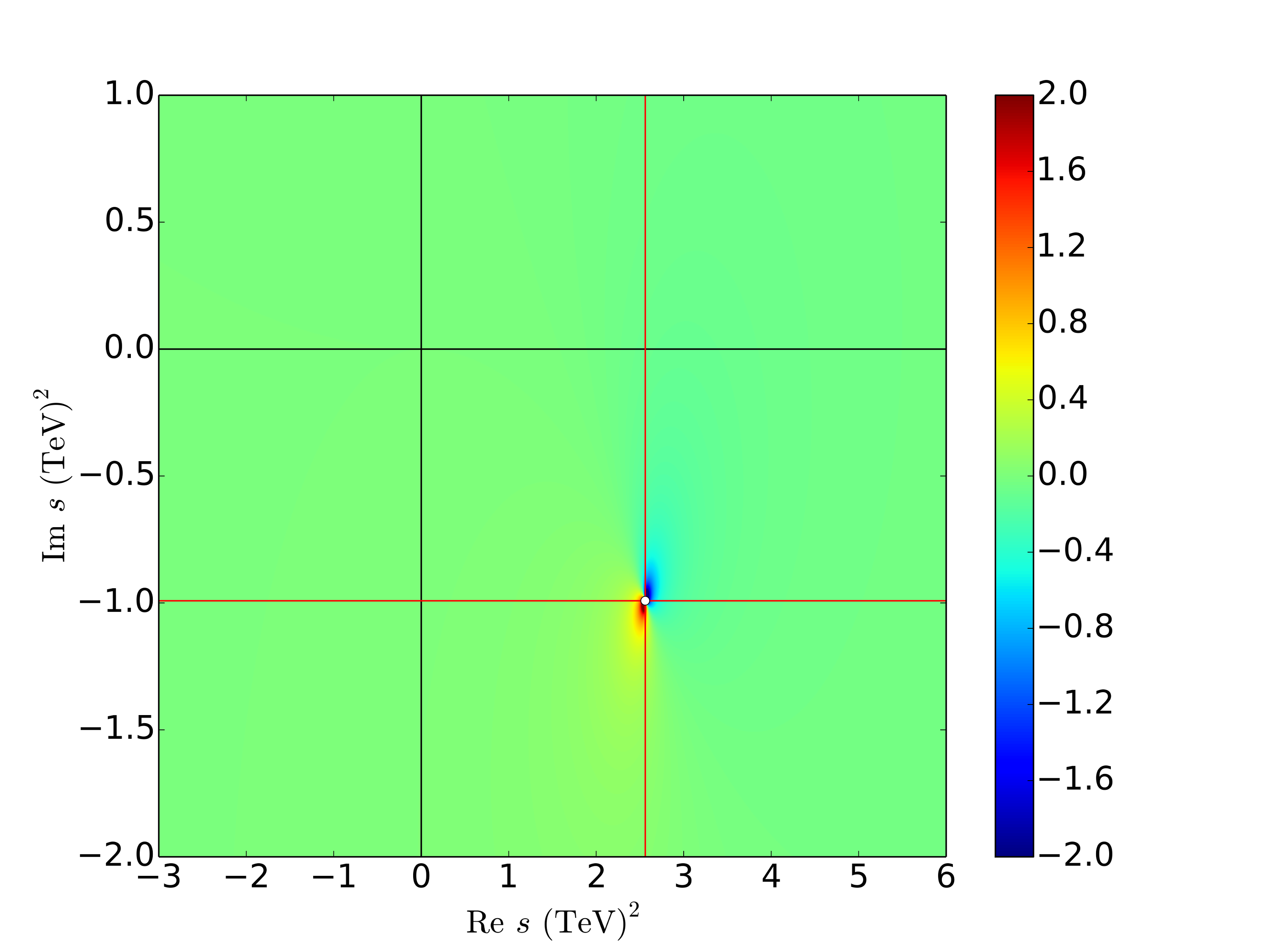}\hfill
\includegraphics[width=.49\textwidth]{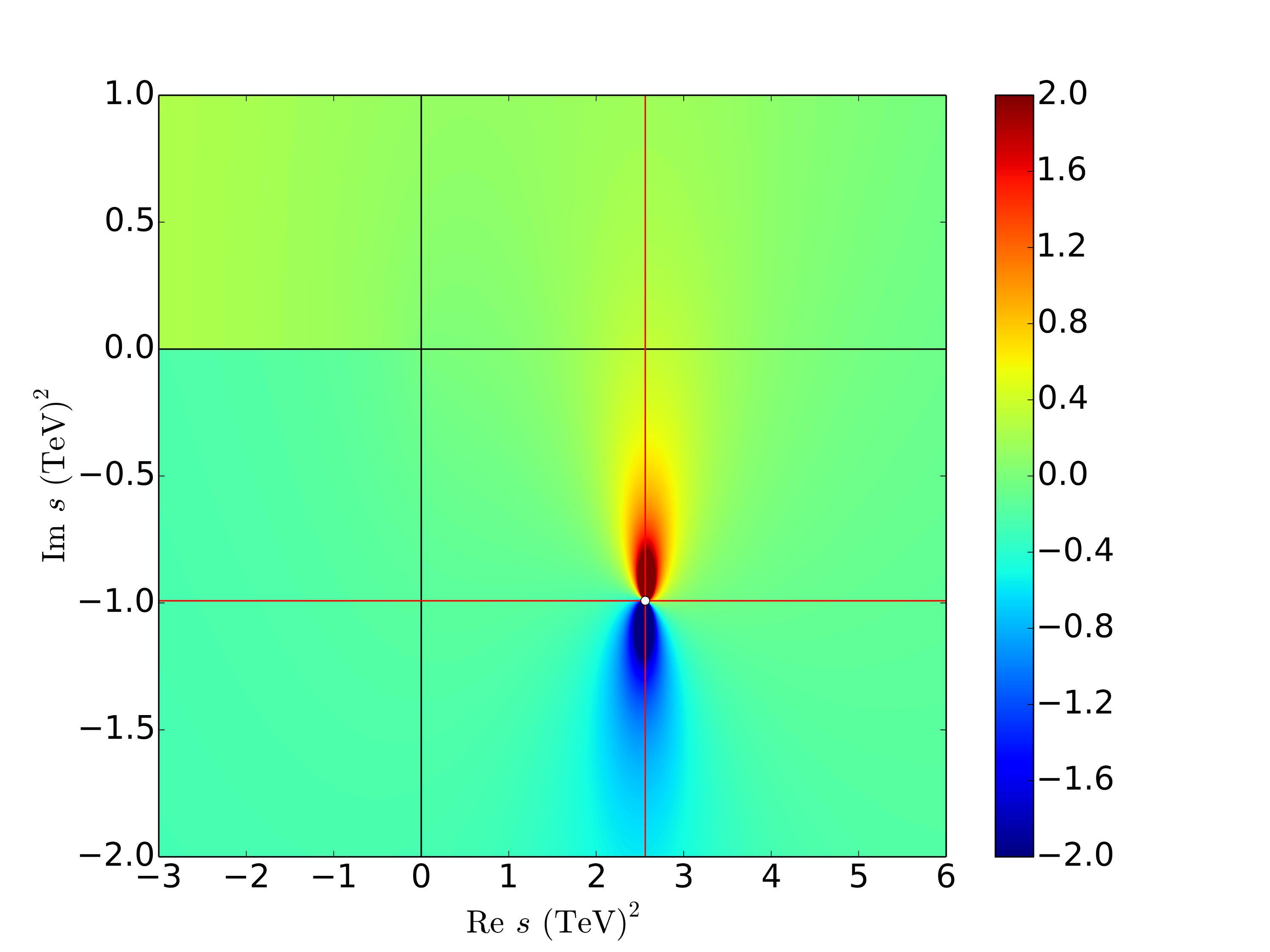}
\caption{From top left, clockwise, $\lvert A_{00}^{\rm IAM}\rvert$, $\lvert Q\rvert$ and $10^2\times\lvert P_{00}\rvert^2$ over the physical region of $s$; and second Riemann sheets of $\Imag A_{00}^{\rm IAM}$, $\Imag Q$ and $\Imag P_{00}$. Same parameters of Fig.~\ref{fig:examples1}.}\label{fig:examples2}
\end{figure}
On Figs.~\ref{fig:examples1} and~\ref{fig:examples2}, a comparison between the $IJ=00$ $A(\omega\omega\to\omega\omega)$, $Q(\omega\omega\to t\bar t)$ and $P(\omega\omega\to\gamma\gamma)$ can be found. We have used $\mu=3\,{\rm TeV}$ (see matrix elements in Refs.~\cite{Delgado:2015kxa,Delgado:2016rtd,Castillo:2016erh}), $M_t=175\,{\rm GeV}$, $a^2=b=0.81^2$, $a_4=4\times 10^{-4}$, $g=10^{-3}$ and all the other NLO counterterms set to $0$. This corresponds to the unitarization of the single channel $\omega\omega\to\omega\omega$ plus weak couplings with $t\bar t$ and $\gamma\gamma$.

For computing the position of poles on the second Riemann sheet (resonances) in Fig.~\ref{fig:examples2}, we have used  Cauchy's Theorem via numerical integration, following the procedure that was already explained in our Ref.~\cite{Delgado:2015kxa}.

Note the presence of a pole (broad resonance) on the second Riemann sheet in the same position in all the three channels (elastic scattering $\omega\omega\to\omega\omega$ and couplings with $t\bar t$ and $\gamma\gamma$). This was expected since the pole comes from the inner EWSBS (strongly coupled) dynamics, not from the physics of the (weak) coupling with $\gamma\gamma$ and $t\bar t$ states.

Poles in the second Riemann sheet, if there are any, should come from the unitarization procedures applied to the inner dynamics of the EWSBS (Eq.~\ref{ec:IAM_EWSBS} and other expressions from our Ref.~\cite{Delgado:2015kxa}). Couplings with $\gamma\gamma$ and $t\bar t$ do not give rise to poles in the second Riemann sheet, according to Eq.~(\ref{ec:PERTURB}) (and similar expressions in Refs.~\cite{Delgado:2016rtd,Castillo:2016erh}), since perturbative partial waves (in particular, $P^{(0)}$ and $A^{(0)}$ in Eq.~\ref{ec:PERTURB}) are just polynomial in $s/v^2$. Thus, the parameters associated with $\gamma\gamma$  ($a_1$, $a_2$, $a_3$, $c_\gamma$) and $t\bar t$ ($c_1$, $c_2$, $g_t$ and $g_t'$) should control the $\gamma\gamma$ and $t\bar t$ coupling strength, but not the physics of the inner dynamics of the EWSBS. This is the picture which emerges from the chiral counting of Fig.~\ref{fig:counting}. (However, if the parameters which control the couplings with $\gamma\gamma$ and $t\bar t$ states were unnaturally large, either $M_t/v$ or $\alpha_{\rm EW}$ would be non-perturbative and a full multi-channel unitarization involving $\{\gamma\gamma,t\bar t\}$ should be carried out~\cite{Castillo:2016erh}.)

\section{Conclusions}
We have computed the NLO scattering processes $\{\omega\omega,hh\}\leftrightarrow\{\gamma\gamma, t\bar t\}$, and coupled them to a (hypothetical) strongly interacting EWSBS within the framework of HEFTs ($E<4\pi v\sim 3\,{\rm TeV}
$). The Equivalence Theorem~\cite{Delgado:2014jda,Delgado:2014dxa,Delgado:2015kxa} (which requires $E>M_h,M_W$) is used.

We have implemented all the relevant perturbative matrix elements and unitarization procedures for $\{\omega\omega,hh\}\leftrightarrow\{\omega\omega,hh,\gamma\gamma,t\bar t\}$ processes inside Fortran modules, following our Refs.~\cite{Delgado:2014jda,Delgado:2014dxa,Delgado:2015kxa,Delgado:2016rtd,Castillo:2016erh}. Several unitarization procedures (based on dispersion relations) are used: IAM, N/D, Improve-K matrix and perturbative couplings with $\gamma\gamma$ and $t\bar t$ states.

Our formalism assumes that the inner EWSBS dynamics (in the $W_L^\pm$, $Z_L$ and $h$ sector) is stronger than their electromagnetic coupling with $\gamma$s ($\sim\alpha_{\rm EM}
$) and the electroweak couplings with $t\bar t$ ($\sim M_t/v$). Hence, any new resonance (within the range $M_W,M_h<E<3\,{\rm TeV}$) in $\{\omega\omega,hh\}\leftrightarrow\{\gamma\gamma,t\bar t\}$ channels would come from the inner EWSBS dynamics. This assumption could be broken for unnaturally large values of the NLO parameters related with $\gamma\gamma$ and $t\bar t$ couplings. Such a situation would require the usage of a full coupled channel unitarization, including strongly $\gamma\gamma$ or $t\bar t$ rescattering.

Even in the case that there are no new resonances, our scattering amplitudes can be a useful tool to parametrize separations from the SM in the regime $M_h,M_W < E < 3\,{\rm TeV}$. Currently, HEFTs themselves are being widely used in this way by CERN collaborations~\cite{YellowReport}.

Finally, we are working within an expanded collaboration in the computation of simple estimates for collider cross-sections of typical resonances, and even for releasing a Monte Carlo module for the experimentalists.


\section*{Acknowledgements}
We thank very useful conversation and suggestions from A. Castillo, Antonio Dobado, J.J. Sanz-Cillero, D. Espriu, and M.J. Herrero. We thank prof.~Felipe J. Llanes-Estrada for reading the 1st version of the manuscript. Work supported by Spanish grants MINECO:FPA2011-27853-C02-01, MINECO:FPA2014-53375-C2-1-P and by BES-2012-056054.


\end{document}